\documentclass[12pt]{article}

\usepackage{epsfig}
\usepackage{amssymb,amsmath}
\usepackage{graphicx, caption}
\usepackage{epstopdf}
\usepackage{color}
\usepackage{setspace}
\newcommand{\bea}{\begin{eqnarray}}
\newcommand{\eea}{\end{eqnarray}}

 \makeatletter
\def\alt{\mathrel{\mathpalette\gl@align<}}
\def\agt{\mathrel{\mathpalette\gl@align>}}
\def\gl@align#1#2{\lower.6ex\vbox{\baselineskip\z@skip\lineskip\z@
\ialign{$\m@th#1\hfil##\hfil$\crcr#2\crcr\sim\crcr}}} \makeatother

\begin{document}
%\begin{flushright}
%BA-13-XX \\
%\end{flushright}
%
%\vspace*{1.0cm}

\begin{center}
\baselineskip 20pt {\Large\bf
125 GeV Higgs boson mass from 5D gauge-Higgs unification
}
\vspace{1cm}

{\large
Jason Carson \footnote{E-mail:jccarson1@crimson.ua.edu}
and Nobuchika Okada \footnote{ E-mail:okadan@ua.edu}
} \vspace{.5cm}

{\baselineskip 20pt \it
Department of Physics and Astronomy,\\
University of Alabama, Tuscaloosa, Alabama 35487, USA }
\vspace{.5cm}

\vspace{1.5cm} {\bf Abstract}
\end{center}
In the context of a simple gauge-Higgs unification (GHU) scenario  
   based on the gauge group SU(3)$\times$U(1)$^\prime$  in a 5-dimensional flat space-time, 
   we investigate a possibility to reproduce the observed Higgs boson mass of around 125 GeV.  
We introduce bulk fermion multiplets with a bulk mass and a (half) periodic boundary condition.  
In our analysis, we adopt a low energy effective theoretical approach of the GHU scenario, 
   where the running Higgs quartic coupling is required to vanish at the compactification scale. 
Under this ``gauge-Higgs condition," we investigate the renormalization group evolution
   of the Higgs quartic coupling and find a relation between the bulk mass and the compactification scale 
   so as to reproduce the 125 GeV Higgs boson mass. 
Through quantum corrections at the one-loop level, 
   the bulk fermions contribute to the Higgs boson production and decay processes 
   and deviate the Higgs boson signal strengths at the Large Hadron Collider (LHC) experiments     
   from the Standard Model (SM) predictions. 
Employing the current experimental data which show the the Higgs boson signal strengths 
   for a variety of Higgs decay modes are consistent with the SM predictions,  
   we obtain lower mass bounds on the lightest mode of the bulk fermions 
   to be around 1 TeV.   
\thispagestyle{empty}

%\bigskip
\newpage

\addtocounter{page}{-1}

%%%%%%%%%%%%%%%%%%%%%%%%%%
%\baselineskip 30pt
% Main body
%%%%%%%%%%%%%%%%%%%%%%%%%%
\baselineskip 18pt
%%%%%%%%%%%%%%%%%%%%%%%%%%

%%%%%%%%%%%%%%%%%%%%%%
\section{Introduction}
%%%%%%%%%%%%%%%%%%%%%%
\label{sec:1}

The discovery of the Standard Model (SM)  Higgs boson 
  by the ATLAS~\cite{ATLAS} and the CMS~\cite{CMS} collaborations 
  at the Large Hadron Collider (LHC) is a milestone in the history of particle physics,   
  and the experimental confirmation of the SM Higgs boson properties has just begun. 
A combined analysis by the ATLAS and the CMS collaborations \cite{ATL_CMS_Hmass} 
  has determined the Higgs boson mass very precisely as 
  $m_H=125.09 \pm 0.21({\rm stat.}) \pm 0.11({\rm syst.})$ GeV.
It also has been shown that by combined ATLAS and CMS measurements \cite{ATL_CMS_Hproperties} 
    that the Higgs boson production and decay rates are consistent with the SM predictions.  
Although the current LHC data include less indications for the direct productions of new particles,  
    the Higgs boson can be a portal to reveal new physics beyond the SM 
    through more precise measurements of the Higgs boson properties 
    and their possible deviations from the SM predictions.

The observed Higgs boson mass of around 125 GeV indicates that the electroweak vacuum 
   is unstable \cite{RGErun}, since the running SM Higgs quartic coupling becomes negative 
   at the energy around $10^{10}$ GeV, for the top quark pole mass $M_t=173.34 \pm 0.76$ 
   from the combined measurements by the Tevatron and the LHC experiments \cite{MtCombine}.\footnote{
The stability of SM Higgs potential is very sensitive to the input value of top quark pole mass, and we
need more precise measurements for it \cite{VacStb2}.
}
This electroweak vacuum instability may not be a problem, since the lifetime of the electroweak vacuum is 
   estimated to be much longer than the age of the universe (meta-stability bound) \cite{meta}. 
However, there are a few discussions which suggest us to take the instability problem seriously: 
the stability of the Minkowski vacuum has been suggested in terms of consistent quantum 
   field theory \cite{Dvali}, and it is unclear if the discussion of the meltability is quantum theoretically consistent or not 
   (see also \cite{counter_discussion}). 
Another discussion is based on inflationary universe.   
If the Higgs potential has a true anti-de Sitter minimum far away from the electroweak minimum, 
   vacuum fluctuations of the Higgs field in de Sitter space during inflation 
   can push the Higgs field to the unwanted anti-de Sitter vacuum \cite{HiggsVacuum_Inflation}. 
Thus, the electroweak vacuum instability might be a serious problem in particle physics and cosmology. 
If this is the case, we need to extend the SM to avoid the running Higgs quartic coupling 
   from turning negative at a high energy. 
For example, we may introduce new physics provided by type II \cite{typeII} and type III \cite{typeIII} 
   seesaw mechanisms which are often invoked  in understanding the solar and atmospheric neutrino oscillations, 
   where the running Higgs quartic coupling remains positive in the presence of new particles 
   \cite{instability_typeII, instability_typeIII, instability_typeIII_GHU}.

In this paper, we consider a novel interpretation for the electroweak vacuum instability problem 
   in terms of  the gauge-Higgs unification (GHU) scenario \cite{GH}, 
   which is one of interesting new physics models beyond the SM.  
In the GHU scenario, the SM Higgs doublet is identified as an extra spatial component 
   of the gauge field in higher dimensional theory, and the higher-dimensional gauge invariance 
   forbids the quadratic divergence in the self-energy corrections of the Higgs doublet 
   in the SM \cite{GH_corrections}.  
As a result, the gauge hierarchy problem can be solved.  
In this paper, we focus on a simple GHU scenario in the flat 5-dimensional space-time  
  and assume that the SM is realized as a low energy effective theory of the scenario.  
It has been pointed out \cite{GHcondition} that in this case 
  the running SM Higgs quartic coupling ($\lambda(\mu)$) must satisfy 
  a special boundary condition (gauge-Higgs condition), 
  namely, $\lambda(M_{\rm KK})=0$, where the $M_{\rm KK}$ is the compactification scale 
  of  the 5th dimension (Kaluza-Klein mass).  
This condition has been derived in Ref.~\cite{GHcondition} as a renormalization condition 
  for the effective Higgs quartic coupling by using an explicit formula 
  of the effective Higgs potential calculated in a simple GHU model. 
Since the SM Higgs doublet field is provided as the 5th component of 
  the 5-dimensional  gauge field, there is no Higgs potential at tree level in the GHU scenario.   
The Higgs potential is radiatively generated at  low energies 
  with the breaking of the original gauge symmetry down to the SM gauge group 
  by a certain boundary condition under a 5th-dimensional coordinate transformation. 
Therefore, in the effective theoretical point of view, we expect that once the original gauge symmetry 
  gets restored at some high energy, the Higgs potential must vanish. 
This is nothing but the gauge-Higgs condition. 
Note that the gauge-Higgs condition leads to a new interpretation 
   for the electroweak vacuum instability problem in the SM, that is, 
   the energy at which $\lambda(\mu)=0$ is nothing but the compactification scale 
   and the 5-dimensional GHU scenario takes place there.

The gauge-Higgs condition is a powerful tool to calculate the Higgs boson mass 
   irrespectively of GHU model details.  
The Higgs boson mass is easily obtained by solving the renormalization group (RG) equation 
   of the Higgs quartic coupling by imposing the gauge-Higgs condition 
   at a given compactification scale, when the particle contents and the mass spectrum 
   of the low energy effective theory are defined below the compactification scale.  
Before the discovery of the SM Higgs boson, 
   the gauge-Higgs condition was utilized to predict the SM Higgs boson mass  
   as a function of the compactification scale, assuming the SM particle contents only \cite{GOS}.  
After the Higgs boson discovery, the 125 GeV Higgs boson mass indicates 
   that the compactification scale lies around $10^{10}$ GeV  \cite{instability_typeIII_GHU, GOS2}.  
Unfortunately, such a compactification scale is too high to be accessible 
   to any ongoing and planned experiments.

Since the Higgs self-energy induced through the Kaluza-Klein (KK) modes of the SM particles 
  (plus some extra matters in the bulk) is proportional to $M_{\rm KK}^2$,  
  the compactification scale around the TeV is desired to solve the gauge hierarchy problem. 
With only the SM particle contents, the GHU scenario with the TeV compactification scale predicts 
  the Higgs boson mass to be too small, $m_H < 100$ GeV.  
In order to realize the 125 GeV mass, we need to introduce extra fermions in the bulk. 
In other words, the observed Higgs boson mass implies the existence of exotic fermions 
  in the context of the GHU scenario. 
In a simple GHU model based on the SU(3)$\times$U(1)$^\prime$ gauge group,  
  the Higgs boson mass was calculated in the presence of some bulk fermion multiplets 
  such as {\bf 10} and {\bf 15} representations under the SU(3) \cite{MO2}.  
It has been shown that the 125 GeV Higgs boson mass can be realized 
   for the TeV compactification scale.  
The contributions of the bulk fermions to the Higgs boson production and decay processes 
   have also been investigated in \cite{MO2}, and the lower bounds on the exotic fermion masses 
   have been obtained from the current LHC data.

The purpose of the present paper is to perform detailed analysis 
   for the GHU model in \cite{MO2} and obtain a more accurate bulk fermion 
   mass spectrum to reproduce the 125 GeV Higgs boson mass.  
In \cite{MO2}, the Higgs boson mass is calculated by solving the RG equation 
   of the Higgs quartic coupling at the leading-log approximation, 
   and no runnings of the gauge couplings and Yukawa couplings have been taken into account. 
Although this analysis would be good enough to estimate the order of the exotic fermion masses,  
   the resultant mass spectrum is not sufficiently accurate 
   in order to discuss the experimental search for the exotic fermions,  
   since the running gauge and Yukawa couplings are expected to be changing a lot 
   in the presence of such a higher-order representation fermions.  
In this paper, we will find that our resultant bulk fermion masses to reproduce the 125 GeV Higgs boson mass
   are quite different from those previously obtained by the rough estimates.

The plan of this paper is as follows. 
In the next section, we introduce a simple GHU model 
   based on the gauge group SU(3)$\times$U(1)$^\prime$ \cite{SSS, CCP} in a 5-dimensional flat space-time 
  with an orbifold $S^1/Z_2$ compactification to the 5th spacial dimension. 
As an example, we introduce bulk fermions in the representations of ${\bf 6}$ and ${\bf 10}$ 
  under the bulk SU(3) gauge group, for which a (half) periodic boundary condition is imposed.
In this context, we evaluate the Higgs boson mass 
  by solving the RG equations with the gauge-Higgs condition 
  and identify the model parameter region to reproduce the observed Higgs boson mass of 125 GeV. 
In Sec.~3, we study effects of the bulk fermions 
  to the Higgs boson production and decay processes at the LHC, 
  and derive a lower mass bound on the lightest bulk fermion from the current LHC data. 
Sec.~4 is devoted to conclusions.

%%%%%%%%%%%%%%%%%%%%%%%%%%%%%%%%%%%%%%%%%%%%%%%%
\section{Higgs boson mass with the gauge-Higgs condition}
%%%%%%%%%%%%%%%%%%%%%%%%%%%%%%%%%%%%%%%%%%%%%%%%
Let us consider a simple GHU model based on the gauge group 
  SU(3)$\times$U(1)$^\prime$ in a 5-dimensional flat space-time 
  with an orbifolding of the 5th dimension on $S^1/Z_2$ with a radius $R_c$ of $S^1$. 
The extra U(1)$^\prime$ symmetry works to yield the correct weak mixing angle, 
  and the SM U(1)$_Y$ gauge boson is given by a linear combination 
  between the gauge bosons of the U(1)$^\prime$ and the U(1) subgroup in SU(3) \cite{SSS}. 
One may think that the U(1)$_X$ gauge boson which is orthogonal to the hypercharge U(1)$_Y$ 
  also has a zero mode. 
However, the U(1)$_X$ symmetry is anomalous in general and broken at the cutoff scale 
  and hence, the U(1)$_X$ gauge boson has a mass of order of the cutoff scale \cite{SSS}. 
As a result, zero-mode vector bosons in the model are only the SM gauge fields.  
In this paper, we employ the effective theoretical approach developed in Ref.~\cite{GHcondition}
  and evaluate the Higgs boson mass with the gauge-Higgs condition. 
In this way, we do not discuss how to provide a complete set of bulk fermions  
  whose zero-modes correspond to the SM fermions. 
Among lots of possibilities, we may refer the proposal in Ref.~\cite{CCP}, 
  where the SM fermions are provided certain SU(3) representations 
  with a suitable U(1)$^\prime$ charge to yield  the correct hypercharges.  
In evaluating the Higgs boson mass, what we need is to define the particle contents   
  for particles lighter than the compactification scale.

The boundary conditions should be suitably assigned to reproduce the SM fields as the zero modes. 
While a periodic boundary condition corresponding to $S^1$ 
  is taken for all of the bulk SM fields, 
  the $Z_2$ parity is assigned for gauge fields and fermions 
  in the representation ${\cal R}$ by using the parity matrix $P={\rm diag}(-,-,+)$ as
\bea
A_\mu (-y) = P^\dag A_\mu(y) P, \quad A_y(-y) =- P^\dag A_y (y) P,  \quad 
\psi(-y) = {\cal R}(P)\psi(y) 
\label{parity}
\eea 
  where the subscripts $\mu$ ($y$) denotes the four (the fifth) dimensional component.
With this choice of parities, the SU(3) gauge symmetry is explicitly broken to SU(2)$\times$U(1). 
The hypercharge U(1)$_Y$ is realized as a linear combination of U(1) and U(1)$^\prime$ in this setup.

With the above parity assignment, off-diagonal blocks in $A_y$ have zero modes, 
   which is identified as the SM Higgs doublet ($H$) such as 
\bea   
A_y^{(0)} = \frac{1}{\sqrt{2}}
\left(
\begin{array}{cc}
0 & H \\
H^\dag & 0 \\
\end{array}
\right). 
\eea
The KK modes of $A_y$ are eaten by KK modes of the SM gauge bosons 
 and enjoy their longitudinal degrees of freedom just like the usual Higgs mechanism.

The parity assignment also provides the SM fermions as massless modes, 
  but in general the massless modes include exotic fermions.   
In order to make such exotic fermions massive, we may introduce 
  brane localized fermions with conjugate SU(2)$\times$U(1)$_Y$ charges 
  and an opposite chirality to the exotic fermions and then write mass terms 
  among the exotic fermions on the orbifold fixed points. 
In the GHU scenario, the Yukawa interaction is unified with the electroweak gauge interaction, 
   so that the SM fermions naturally have the mass of the order of the $W$-boson mass 
   after the electroweak symmetry breaking. 
This feature is good only for the top quark, while most of the SM fermions are much lighter 
   than the weak boson.     
To realize light SM fermion masses, one may introduce a $Z_2$-parity odd bulk mass terms for the bulk SM fermions. 
In the presence of the parity-odd bulk mass, zero mode fermion wave functions 
  with opposite chirality are localized towards the opposite orbifold fixed points 
  and as a result, their effective 4-dimensional Yukawa coupling is exponentially 
  suppressed by the overlap integral of the wave functions. 
In this way, we assume that all exotic fermion zero modes become very heavy 
  and realistic SM fermion mass matrices are achieved by adjusting the bulk mass parameters. 
For more details towards constructing a realistic GHU scenario, see, for example, Refs.~\cite{SSS, CCP}.

Let us now investigate the way to reproduce the Higgs boson mass of around 125 GeV 
   in this 5-dimensional GHU model. 
It is a highly non-trivial task to propose a realistic GHU scenario 
   and calculate the Higgs boson mass in the context. 
However, in our effective theory approach, the Higgs boson mass 
  is easily calculated from the RG evolution of the Higgs quartic coupling 
  with the gauge-Higgs condition at the compactification scale,  
  assuming the electroweak symmetry breaking is correctly achieved. 
In order to reproduce the 125 GeV Higgs boson mass for $M_{\rm KK} \ll 10^{10}$ GeV, 
  we need to introduce a new fermion in the bulk. 
In this paper, we introduce color singlet/triplet, ${\bf 6}$ and ${\bf 10}$-plet bulk fermions
  of the bulk SU(3) gauge symmetry with U(1)$^\prime$ charge $Q$ as an example. 
We impose a (half) periodic boundary condition on the bulk fermions, 
   $\psi(y+2\pi R_c) = \psi(y)$ ($\psi(y+2\pi R_c) = -\psi(y)$).  
To avoid massless states in the periodic bulk fermions, 
  we introduce $N_f$ pairs of the bulk fermion multiplets with opposite parities 
  and a $Z_2$-parity even bulk mass term between each pair of the bulk fermions.  
In the same way, we introduce $N_f^{\rm HP}$ pairs of half-periodic fermions 
  with the $Z_2$-parity even bulk mass term, when we consider half-periodic bulk fermions.\footnote{
Since no massless mode exists for the half-periodic bulk fermions, 
  the bulk mass term is unnecessary for them. 
However, the bulk mass parameter along with the other free parameter $R_c$ 
   simplifies our analysis to reproduce the 125 GeV Higgs boson mass.  
The case with no bulk mass corresponds to $m_0=M_{\rm KK}/2$ (for notations, see below Eq.~(\ref{6KKspectrum})).  
In this case, we cannot reproduce the 125 GeV Higgs boson mass, as we can see from Fig.~\ref{Fig:MkkVSm0-10}. 
}

We begin with the ${\bf 6}$-plet of the bulk SU(3) gauge symmetry,  
  which is decomposed into the representations
  under the SU(2)$\times$U(1) subgroup as  
\bea
 {\bf 6} = {\bf 1}_{-2/3} \oplus {\bf 2}_{-1/6} \oplus {\bf 3}_{1/3},  
\label{6deco}
\eea
   where the numbers in the subscripts denote the U(1) charges. 
For these multiplets, the bulk SU(3) gauge interaction leads to the Yukawa interaction of the form, 
\bea 
   {\cal L} \supset - Y_S {\overline D} H S - Y_D {\overline D} T H^\dagger,   
\eea
where $S$, $D$ and $T$ stand for the singlet, doublet and triplet fields 
   in the decomposition of Eq.~(\ref{6deco}), and $Y_S$ and $Y_D$ are Yukawa couplings.  
Because of the unification of the gauge and Yukawa interactions, 
   $Y_S = Y_D = - i g_2$ at the compactification scale, where $g_2$ is the SM SU(2) gauge coupling. 
In solving RG equations, this condition is also imposed as the boundary condition at the compactification scale.   
After the electroweak symmetry breaking  the KK mass spectrum is found as follows: 
\bea
&& 
\left( m_{n,-2/3}^{(\pm)} \right)^2 
 = \left( m_n \pm 2 m_W \right)^2 +M^2,~~ 
   m_{n}^2 + M^2, \nonumber \\ 
&& 
\left( m_{n,+1/3}^{(\pm)} \right)^2 
 = \left( m_{n} \pm m_W \right)^2 +M^2, \nonumber \\ 
&& 
\left( m_{n,+4/3}^{(\pm)} \right)^2 
 = m_{n}^2 +M^2, 
\label{6KKspectrum}
\eea  
 where the numbers in the subscripts denote the ``electric charges"\footnote{
  Here ``electric charges" mean by electric charges of SU(2)$\times$U(1)$\subset$SU(3). 
  A true electric charge of each KK mode is given 
  by a sum of the ``electric charge" and U(1)$^\prime$ charge $Q$.
 }  
  of the corresponding KK mode fermions, 
  $m_n= n M_{\rm KK}$ with $n=0,1,2,\cdots$, $M_{\rm KK} \equiv 1/R_c$, 
  $m_W=g_2 v/2$ with $v=246$ GeV, and $M$ is a bulk mass.    
For simplicity, we use a common bulk mass $M$ for the $N_f$ pairs.  
When a half-periodic boundary condition is imposed on the bulk fermion, 
   the KK mass spectrum are obtained by replacing $n$ to $n+1/2$.

In the same way, we decompose the ${\bf 10}$-plet as 
\bea
 {\bf 10} = {\bf 1}_{-1} \oplus {\bf 2}_{-1/2} 
 \oplus {\bf 3}_{0} \oplus {\bf 4}_{1/2}.   
\label{10deco}
\eea
For these SM multiplets, the bulk SU(3) gauge interaction leads to the Yukawa interaction of the form, 
\bea 
   {\cal L} \supset - Y_S {\overline D} H S - Y_D {\overline D} T H^\dagger - Y_T {\overline F} T H,   
\eea
where $S$, $D$, $T$ and $F$ stand for the singlet, doublet, triplet and quartet fields 
   in the decomposition of Eq.~(\ref{10deco}), and $Y_S$, $Y_D$ and $Y_T$ are Yukawa couplings.  
Because of the unification of the gauge and Yukawa interactions, 
   $Y_S = Y_T = - i \sqrt{3/2} \; g_2$ and $Y_D=-i \sqrt{2} \; g_2$ at the compactification scale. 
These conditions are imposed as the boundary condition at the compactification scale in our RG analysis. 
The KK mass spectrum after the electroweak symmetry breaking is found as 
\bea
&& 
\left( m_{n,-1}^{(\pm)} \right)^2 
 = \left( m_{n} \pm 3 m_W \right)^2 +M^2,~~ 
   \left( m_{n} \pm   m_W \right)^2 +M^2, \nonumber \\ 
&& 
\left( m_{n,0}^{(\pm)} \right)^2 
 = \left( m_{n} \pm 2 m_W \right)^2 +M^2,~~ 
   m_{n}^2 + M^2, \nonumber \\ 
&& 
\left( m_{n,+1}^{(\pm)} \right)^2 
 = \left( m_{n} \pm m_W \right)^2 +M^2, \nonumber \\ 
&& 
\left( m_{n,+2}^{(\pm)} \right)^2 = m_{n}^2 + M^2. 
\label{10KKspectrum}
\eea

Although the U(1)$^\prime$ charge $Q$ is a free parameter of the model, 
  we have phenomenologically favored values for it from the following discussion. 
As discussed in Ref.~\cite{GHDM} (see also Ref.~\cite{GHDM2}),
   the lightest KK mode of a half-periodic bulk fermion, independently of the background metric, 
   is stable in the effective 4-dimensional theory due to an accidental $Z_2$ discrete symmetry. 
If the half-periodic bulk fermion is color-singlet, it is a good candidate for the cosmological dark matter. 
Even for the periodic bulk fermion, we are allowed to introduce an odd-parity 
  while all the SM particles are even under the parity, 
  in order to ensure the stability of the lightest KK mode.    
Thus, it is reasonable to assign the U(1)$^\prime$ charge $Q$ to make the lightest KK mode electrically neutral. 
Since the electric charge is given by the sum of the charge of the U(1) subgroup in the bulk SU(3) and $Q$, 
   we may choose $Q=2/3$ ($Q=1$) for a color-singlet, ${\bf 6}$-plet (${\bf 10}$-plet) bulk fermion. 
However, a colored stable particle is cosmologically disfavored.  
For a color-triplet bulk fermion, we may introduce a mixing between the lightest colored KK fermion 
  and a SM quark on the brane, so that the lightest KK fermion can decay to the SM quarks. 
There are two choices for the U(1)$^\prime$ charge to make the electric charge of the lightest KK mode 
 to be $-1/3$ or $2/3$ for realizing a mixing with either the SM down-type quarks or up-type quarks. 
For the ${\bf 6}$-plet case, we may choose $Q=1/3$ or $4/3$, 
   while $Q=2/3$ or $5/3$ for the ${\bf 10}$-plet case.

Let us now analyze the RG equations. 
In our analysis, we neglect the KK mode mass splitting by the electroweak symmetry breaking 
  and set the lightest fermion mass as $m_0= M$. 
When we impose a half periodic boundary condition for the bulk fermions,   
\bea 
 m_0= \frac{1}{2} M_{\rm KK} \sqrt{1+4 c_B^2}
\eea 
where $c_B \equiv M/M_{\rm KK}$.
For renormalization scale $\mu < m_0$, 
   the bulk fermions are decoupled, and we employ the SM RG equations at the two-loop level~\cite{RGErun}.  
For the three SM gauge couplings $g_i$ ($i=1,2,3$), we have 
\bea
 \frac{d g_i}{d \ln \mu} =
 \frac{b_i}{16 \pi^2} g_i^3 +\frac{g_i^3}{(16\pi^2)^2}
  \left( \sum_{j=1}^3 B_{ij}g_j^2 - C_i y_t^2   \right),
\eea
 where the first and second terms in the right hand side are the beta functions 
 at the one-loop and the two-loop levels, respectively, with the coefficients,  
\bea
b_i = \left(\frac{41}{10},-\frac{19}{6},-7\right),~~~~
 { B_{ij}} =
 \left(
  \begin{array}{ccc}
  \frac{199}{50}& \frac{27}{10}&\frac{44}{5}\\
 \frac{9}{10} & \frac{35}{6}&12 \\
 \frac{11}{10}&\frac{9}{2}&-26
  \end{array}
 \right),  ~~~~
C_i=\left( \frac{17}{10}, \frac{3}{2}, 2 \right). 
\eea 
For contributions from the SM Yukawa coupling to the beta function at the two-loop level, 
 we have considered only the top Yukawa coupling ($y_t$). 
The RG equation for the top Yukawa coupling is given by
\bea 
 \frac{d y_t}{d \ln \mu}
 = y_t  \left(
 \frac{1}{16 \pi^2} \beta_t^{(1)} + \frac{1}{(16 \pi^2)^2} \beta_t^{(2)}
 \right), 
\eea
where the one-loop contribution is
\bea
 \beta_t^{(1)} =  \frac{9}{2} y_t^2 -
  \left(
    \frac{17}{20} g_1^2 + \frac{9}{4} g_2^2 + 8 g_3^2
  \right) ,
\eea
while the two-loop contribution is given by 
\bea
\beta_t^{(2)} &=&
 -12 y_t^4 +   \left(
    \frac{393}{80} g_1^2 + \frac{225}{16} g_2^2  + 36 g_3^2
   \right)  y_t^2  \nonumber \\
 &&+ \frac{1187}{600} g_1^4 - \frac{9}{20} g_1^2 g_2^2 +
  \frac{19}{15} g_1^2 g_3^2
  - \frac{23}{4}  g_2^4  + 9  g_2^2 g_3^2  - 108 g_3^4 \nonumber \\
 &&+ \frac{3}{2} \lambda^2 - 6 \lambda y_t^2 .
\eea
The RG equation for the quartic Higgs coupling is given by 
\bea
\frac{d \lambda}{d \ln \mu}
 =   \frac{1}{16 \pi^2} \beta_\lambda^{(1)}
   + \frac{1}{(16 \pi^2)^2}  \beta_\lambda^{(2)},
\eea
with the one-loop beta function, 
\bea
 \beta_\lambda^{(1)} &=& 12 \lambda^2 -
 \left(  \frac{9}{5} g_1^2+ 9 g_2^2  \right) \lambda
 + \frac{9}{4}  \left(
 \frac{3}{25} g_1^4 + \frac{2}{5} g_1^2 g_2^2 +g_2^4
 \right) + 12 y_t^2 \lambda  - 12 y_t^4 ,
\eea
and the two-loop beta function, 
\bea
  \beta_\lambda^{(2)} &=&
 -78 \lambda^3  + 18 \left( \frac{3}{5} g_1^2 + 3 g_2^2 \right) \lambda^2
 - \left( \frac{73}{8} g_2^4  - \frac{117}{20} g_1^2 g_2^2
 - \frac{1887}{200} g_1^4  \right) \lambda - 3 \lambda y_t^4
 \nonumber \\
 &&+ \frac{305}{8} g_2^6 - \frac{289}{40} g_1^2 g_2^4
 - \frac{1677}{200} g_1^4 g_2^2 - \frac{3411}{1000} g_1^6
 - 64 g_3^2 y_t^4 - \frac{16}{5} g_1^2 y_t^4
 - \frac{9}{2} g_2^4 y_t^2
 \nonumber \\
 && + 10 \lambda \left(
  \frac{17}{20} g_1^2 + \frac{9}{4} g_2^2 + 8 g_3^2 \right) y_t^2
 -\frac{3}{5} g_1^2 \left(\frac{57}{10} g_1^2 - 21 g_2^2 \right)
  y_t^2  - 72 \lambda^2 y_t^2  + 60 y_t^6.
\eea
In solving the RG equations, we use the boundary conditions at the top quark pole mass ($M_t$)
  given in Ref.~\cite{RGErun}: 
\bea 
g_1(M_t)&=&\sqrt{\frac{5}{3}} \left( 0.35761 + 0.00011 (M_t - 173.10) 
   -0.00021  \left( \frac{m_W-80.384}{0.014} \right) \right),  
\nonumber \\
g_2(M_t)&=& 0.64822 + 0.00004 (M_t - 173.10) + 0.00011  \left( \frac{m_W-80.384}{0.014} \right) , 
\nonumber \\
g_3(M_t)&=& 1.1666 + 0.00314 \left(  \frac{\alpha_s-0.1184}{0.0007}   \right) ,
\nonumber \\
y_t(M_t) &=& 0.93558 + 0.0055 (Mt - 173.10) - 0.00042  \left(  \frac{\alpha_s-0.1184}{0.0007}   \right)
  - 0.00042 \left( \frac{m_W-80.384}{0.014} \right) ,
\nonumber \\
\lambda(M_t) &=&  2 (0.12711 + 0.00206 (m_H - 125.66) - 0.00004 (M_t - 173.10) ) .
\eea 
We employ $m_W=80.384$ (in GeV), $\alpha_s =0.1184$, 
   $M_t=173.34$ (in GeV)
   from the combined measurements by the Tevatron and the LHC experiments \cite{MtCombine}, 
   and $m_H=125.09$ (in GeV) from the combined analysis by the ATLAS and the CMS  \cite{ATL_CMS_Hmass}.

For the renormalization scale $\mu \geq m_0$, the SM RG equations are modified 
   in the presence of the bulk fermions.  
In this paper, we take only one-loop corrections from the bulk fermions into account.  
For the case with $N_f$ pairs of {\bf 6}-plet periodic fermions  
   (we identify $N_f$ as $N_f=2 N_f^{\rm HP}$ for the case with $N_f^{\rm HP}$ pairs of half-periodic fermions), 
   the beta functions of the SU(2) and U(1)$_Y$ gauge couplings receive new contributions as 
\bea
\Delta b_1= N_f  N_c \left(\frac{2}{3} + \frac{24}{5} Q^2 \right), \; \; \; 
\Delta b_2 =\frac{10}{3} N_f  N_c,  \; \; \; 
\Delta b_3= 4 N_f  \left( \frac{N_c-1}{2} \right), 
\eea
where $N_c=1$ ($N_c=3)$ when the {\bf 6}-plet bulk fermions are color singlet (triplet). 
The beta functions of the top Yukawa and Higgs quartic couplings are modified as 
\bea 
&& \beta_t^{(1)} \to \beta_t^{(1)}  +y_t N_f N_c \left(2 |Y_S|^2 + 3 |Y_D|^2 \right),  \nonumber \\
&& \beta_{\lambda}^{(1)} \to \beta_{\lambda}^{(1)} +
       N_f  N_c \left[ \lambda \left( 8 |Y_S|^2 + 12 |Y_D|^2 \right) 
       - \left( 8 |Y_S|^4 + 10 |Y_D|^4 +16 |Y_S|^2 |Y_D|^2 \right) \right], 
\label{LamBeta}
\eea
where the Yukawa couplings obey the following RG equations: 
\bea 
16 \pi^2 \frac{d Y_S}{d \ln \mu} &=&
  Y_S \left[ 3 y_t^2 - \left( \frac{9}{20}  g_1^2  + \frac{9}{4} g_2^2 \right) 
  + N_f \left( \frac{4 N_c+3}{2} |Y_S|^2 + \frac{12 N_c+7}{4}  |Y_D|^2 \right) \right. \nonumber \\
  &-& (N_c^2-1) g_3^2 -\left.  \frac{18}{5} \left( \frac{2}{3}- Q\right) \left(\frac{1}{6}-Q \right) g_1^2 \right], \nonumber \\
16 \pi^2 \frac{d Y_D}{d \ln \mu}  &=&
   Y_D  \left[ 3 y_t^2  - \left( \frac{9}{20} g_1^2 + \frac{9}{4} g_2^2 \right)
   + N_f \left( \frac{4 N_c+5}{2}  |Y_S|^2 + \frac{12 N_c+5}{4} |Y_D|^2  \right)  \right. \nonumber \\
   &-& (N_c^2-1) g_3^2 - \left. 6 g_2^2 - \frac{18}{5} \left(\frac{1}{6} - Q\right) \left(\frac{1}{3} -Q \right) g_1^2  \right]. 
\eea

In our RG analysis, we numerically solve the SM RG equations from $M_t$ to $m_0$, 
    at which the solutions connect with the solutions of the RG equations with the bulk fermions. 
For a fixed $m_0$ values, we arrange the input values for $Y_S(m_0)$ and $Y_D(m_0)$ 
   so as to find the numerical solutions which satisfy 
   the gauge-Higgs condition and the unification condition 
   between the gauge and Yukawa couplings at the compactification scale: 
\bea 
  \lambda(M_{\rm KK})=0, \; \; \; Y_S(M_{\rm KK})= Y_D(M_{\rm KK}) = - i g_2 (M_{\rm KK}) . 
\eea

%%%%%%%%%%%%%%%%%%%%%%%%%%%%%%%%%%%%%%%%%%%%%%%%%%%%
% Fig 1
%%%%%%%%%%%%%%%%%%%%%%%%%%%%%%%%%%%%%%%%%%%%%%%%%%%%%
%\begin{figure}[htbp]
\begin{figure}[t]
  \begin{center}
 \includegraphics[width=130mm]{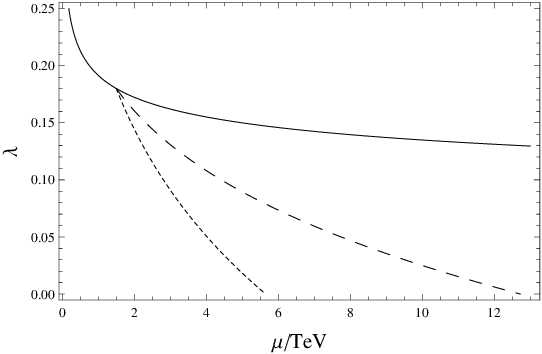}
   \end{center}
\caption{
The RG evolutions of the Higgs quartic coupling, 
  which can reproduce the Higgs boson pole mass of $m_H=125.09$ GeV.  
The solid line denotes the running Higgs quartic coupling in the SM.  
The dashed and dotted lines correspond, respectively, to 
   the ${\bf 6}$-plets 
   for  $N_f=2$, $N_c=1$, $Q=2/3$ and $(M_{\rm KK}, m_0)=(12.7, 1.5)$ TeV,  and 
   the ${\bf 6}$-plet 
   for  $N_f=1$, $N_c=3$, $Q=4/3$ and $(M_{\rm KK}, m_0)=(5.65, 1.5)$ TeV.   
}
  \label{RGE-Q}
\end{figure}
%%%%%%%%%%%%%%%%%%%%%%%%%%%%%%%%%%%%%%%%%%%%%%%%%%%%%

%%%%%%%%%%%%%%%%%
% Fig 2
%%%%%%%%%%%%%%%%%
%\begin{figure}[htbp]
\begin{figure}[t]
  \begin{center}
 \includegraphics[width=130mm]{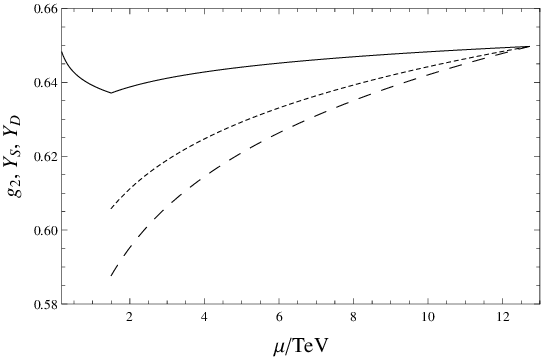}
   \end{center}
\caption{
The RG evolutions of the SM SU(2) gauge coupling (solid line)
   and Yukawa couplings, $|Y_S|$ (dashed line)  and $|Y_D|$ (dotted line),  
   for the case with $N_f=2$ pair of ${\bf 6}$-plet, color singlet bulk fermions with $Q=2/3$.  
We can see that the boundary condition from the unification among the gauge and Yukawa couplings, 
  $|Y_S|=|Y_D|=g_2$, is satisfied at $M_{\rm KK}=12.7$  TeV. 
 }
  \label{RGE_Y}
\end{figure}
%%%%%%%%%%%%%%%%%%%%%%%%%%%%%%%%%%%%%%%%%%%%%%%%%%%%%

The running Higgs quartic coupling to reproduce the Higgs boson pole mass of 
  $m_H=125.09$ GeV is shown in Fig.~\ref{RGE-Q}.    
The solid line denotes the running quartic coupling in the SM, 
  while the dashed (dotted) line corresponds to the result
  for the case with $N_f=2$ ($N_f=1$) pair of ${\bf 6}$-plet, color singlet (triplet) bulk fermions  
  with U(1)$^\prime$ charge $Q=2/3$ ($Q=4/3$).  
For the dashed (dotted) line, we find $M_{\rm KK}=12.7$ ($5.65$) TeV for $m_0=1.5$ TeV, 
   at which the gauge-Higgs condition is satisfied. 
When we trace the dashed and dotted lines from $M_t$ to higher energies 
   we see that the running of the Higgs quartic coupling is drastically altered 
   from the SM one (solid line) due to the contributions from the bulk fermions 
   with $m_0=1.5$ TeV.  
Since the beta function of the Higgs quartic coupling becomes more negative 
  in the presence of the bulk fermions (see Eq.~(\ref{LamBeta})), 
  the running Higgs quartic coupling reaches the compactification scale far below $10^{10}$ GeV. 
We also show in Fig.~\ref{RGE_Y} the RG evolutions of the SM SU(2) gauge coupling (solid line)
   and Yukawa couplings $|Y_S|$ (dashed line)  and $|Y_D|$ (dotted line)  
   for the case with $N_f=2$ pair of ${\bf 6}$-plet, color singlet bulk fermions with $Q=2/3$, 
   corresponding to the dashed line in Fig.~\ref{RGE-Q}. 
We can see that the boundary condition from the unification between the gauge and Yukawa couplings, 
  $|Y_S|=|Y_D|=g_2$, is satisfied at $M_{\rm KK}=12.7$  TeV. 
It is interesting to compare Fig.~\ref{RGE-Q} with Fig.~4 in the first paper in Ref.~\cite{MO2}. 
We can see that the RG evolutions shown these figures are quite different. 
This difference is due to the RG effects of the gauge and Yukawa couplings that are taken into account 
 in our analysis. 
Hence, our resultant bulk fermion masses to reproduce the 125 GeV Higgs boson mass
   are quite different from those obtained in Ref.~\cite{MO2}, 
   as we have mentioned in Sec.~\ref{sec:1}.

%%%%%%%%%%%%%%
%  Fig 3
%%%%%%%%%%%%%%
\begin{figure}[htbp]
  \begin{center}
   \includegraphics[width=80mm]{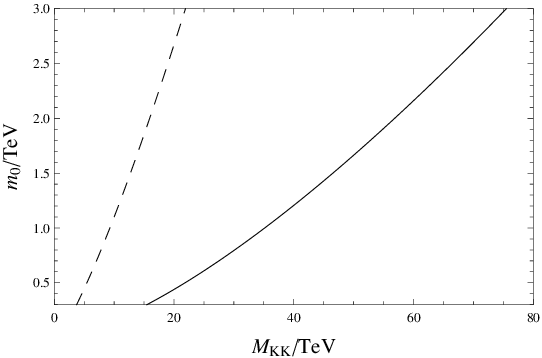}
   \hspace*{5mm}
   \includegraphics[width=80mm]{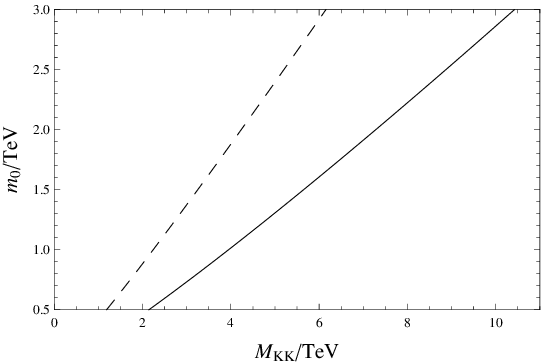}
   \end{center}
\caption{
The relation between $M_{\rm KK}$  and $m_0$ to reproduce the Higgs boson pole mass of 
  $m_H=125.09$ GeV. 
The left panel shows the results for the color singlet, ${\bf 6}$-plet bulk fermions  
   with $N_f=1$ (solid line) and $N_f=2$ (dashed line). 
Here we have taken $Q=2/3$.  
The right panel shows the results for the color triplet, ${\bf 6}$-plet bulk fermions  
   with $N_f=1$ (solid line) and $N_f=2$ (dashed line).  
For the color triplet case, we have taken $Q=4/3$.  
}
  \label{Fig:MkkVSm0}
\end{figure}
%%%%%%%%%%%%%%%%%%%%%%%%%%%%%%%%%%%%%%%%%%%%%%%%%%%%%

Once the lightest fermion mass $m_0$ is fixed, 
   the compactification scale $M_{\rm KK}$ is determined,  
   where the gauge-Higgs condition and the unification condition 
   between the gauge and Yukawa couplings are satisfied.    
The relation between $M_{\rm KK}$ and $m_0$ is depicted in Fig.~\ref{Fig:MkkVSm0}. 
In the left panel we show the relation for the color singlet, ${\bf 6}$-plet bulk fermions  
   with $N_f=1$ (solid line) and $N_f=2$ (dashed line). 
Here we have taken $Q=2/3$ as an example. 
The relations for the color triplet, ${\bf 6}$-plet bulk fermions  
   with $N_f=1$ (solid line) and $N_f=2$ (dashed line), respectively, are shown in the right panel.   
For the color triplet case, we have taken $Q=4/3$.  
As the number of bulk fermions is increasing, the compactification scale for a fixed $m_0$ is decreasing.

%%%%%%%%%%%%%%%%%%%%%%%%%%%%%%%%%%%%%%%%%%%%%%%%%%%%
% Fig 4
%%%%%%%%%%%%%%%%%%%%%%%%%%%%%%%%%%%%%%%%%%%%%%%%%%%%%
%\begin{figure}[htbp]
\begin{figure}[t]
  \begin{center}
 \includegraphics[width=130mm]{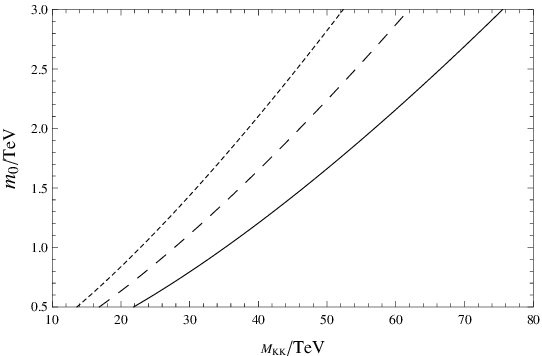}
   \end{center}
\caption{
In the case with the color singlet, ${\bf 6}$-plet bulk fermions ($N_f=1$), 
  the relation between $M_{\rm KK}$  and $m_0$ to reproduce the Higgs boson pole mass of 
  $m_H=125.09$ GeV for various U(1)$^\prime$ charges.   
The solid, dashed and dotted lines corresponds to the results for 
  $Q=0$, $2$ and $-2$, respectively.  
}
  \label{Fig:MkkVSm0-Q}
\end{figure}
%%%%%%%%%%%%%%%%%%%%%%%%%%%%%%%%%%%%%%%%%%%%%%%%%%%%%

For the color singlet, ${\bf 6}$-plet bulk fermions  with $N_f=1$, 
  we show in Fig.~\ref{Fig:MkkVSm0-Q} the relation between $M_{\rm KK}$ and $m_0$ 
  for various U(1)$^\prime$ charges.  
The solid, dashed and dotted lines corresponds to the results for 
  $Q=0$, $2$ and $-2$, respectively.  
We find that the compactification scale for a fixed $m_0$ is decreasing, as $|Q|$ is increasing.

We perform the same analysis for the case with the $N_f$ pairs of the {\bf 10}-plet periodic fermions 
 (we identify $N_f$ as $N_f=2 N_f^{\rm HP}$ for the case with $N_f^{\rm HP}$ pairs of half-periodic fermions).  
For the renormalization scale $\mu \geq m_0$,  the beta functions of 
  the SU(2) and U(1)$_Y$ gauge couplings receive new contributions as 
\bea
\Delta b_1= N_f  N_c \left( 3+ 12 Q^2 \right), \; \; \; 
\Delta b_2 = 10 N_f  N_c,  \; \; \; 
\Delta b_3=\frac{10}{3} N_f  \left( \frac{N_c-1}{2} \right), 
\eea
where $N_c=1$ ($N_c=3)$ when the {\bf 10}-plet bulk fermions are color singlet (triplet). 
The beta functions of the top Yukawa and Higgs quartic couplings are modified as 
\bea 
 \beta_t^{(1)} &\to& \beta_t^{(1)}  +y_t N_f N_c \left(2 |Y_S|^2 + 3 |Y_D|^2 \right),  \nonumber \\
 \beta_{\lambda}^{(1)} &\to& \beta_{\lambda}^{(1)} +
       N_f  N_c \left[ 4 \lambda \left( 2 |Y_S|^2 + 3 |Y_D|^2 +4 |Y_T|^2\right)  \right. \nonumber \\
& & \left. - \left( 8 |Y_S|^4 + 10 |Y_D|^4 + \frac{112}{9} |Y_T|^4 
                 +16 |Y_S|^2 |Y_D|^2 + \frac{64}{3} |Y_D|^2 |Y_T|^2  \right) \right], 
\label{LamBeta-10}
\eea
where the Yukawa couplings obey the following RG equations: 
\bea 
16 \pi^2 \frac{d Y_S}{d \ln \mu} &=&
  Y_S \left[ 3 y_t^2 - \left( \frac{9}{20}  g_1^2  + \frac{9}{4} g_2^2 \right) 
  + N_f \left( \frac{4 N_c+3}{2} |Y_S|^2 + \frac{12 N_c+15}{4}  |Y_D|^2  + 4 N_c |Y_T|^2  \right) \right. \nonumber \\
  &-& (N_c^2-1) g_3^2 -\left.  \frac{18}{5} \left(1- Q\right) \left(\frac{1}{2}-Q \right) g_1^2 \right], \nonumber \\
16 \pi^2 \frac{d Y_D}{d \ln \mu}  &=&
   Y_D  \left[ 3 y_t^2  - \left( \frac{9}{20} g_1^2 + \frac{9}{4} g_2^2 \right)
   + N_f \left( \frac{4 N_c+5}{2}  |Y_S|^2 + \frac{12 N_c+5}{4} |Y_D|^2 + \frac{12 N_c+10}{3} |Y_T|^2 \right)  \right. \nonumber \\
   &-& (N_c^2-1) g_3^2 - \left. 6 g_2^2 - \frac{18}{5} \left(Q-\frac{1}{2} \right) Q \;  g_1^2  \right],  \nonumber \\ 
16 \pi^2 \frac{d Y_T}{d \ln \mu}  &=&
   Y_T  \left[ 3 y_t^2  - \left( \frac{9}{20} g_1^2 + \frac{9}{4} g_2^2 \right)
   + N_f \left( 2 N_c  |Y_S|^2 + \frac{6 N_c+5}{2} |Y_D|^2 + \frac{24 N_c+7}{6} |Y_T|^2 \right)  \right. \nonumber \\
   &-& (N_c^2-1) g_3^2 - \left. 15 g_2^2 - \frac{18}{5} Q \left(Q+\frac{1}{2} \right) g_1^2  \right]. 
\eea

We numerically solve the SM RG equations from $M_t$ to $m_0$, 
    at which the solutions connect with the solutions of the RG equations with the ${\bf 10}$-plet bulk fermions. 
For a fixed $m_0$ values, we arrange the input values for $Y_S(m_0)$, $Y_D(m_0)$ and $Y_T(m_0)$
   so as to find the numerical solutions which satisfy 
   the gauge-Higgs condition and the unification condition 
   between the gauge and Yukawa couplings at the compactification scale: 
\bea 
  \lambda(M_{\rm KK})=0, \; \; \; 
 \sqrt{\frac{2}{3}} Y_S(M_{\rm KK}) = \sqrt{\frac{1}{2}} Y_D(M_{\rm KK})=   \sqrt{\frac{2}{3}} Y_T(M_{\rm KK})= - i g_2 (M_{\rm KK}) . 
\eea      
In Fig.~\ref{Fig:MkkVSm0-10}, we show the relation between $M_{\rm KK}$ and $m_0$.  
The solid and dashed lines denotes the results for the color singlet, ${\bf 10}$-plet bulk fermions 
  with $N_f = 1$ and $N_f = 2$, respectively. 
Here we have taken $Q = 1$.  
The dotted and dash-dotted lines represent the results for the color triplet, ${\bf 10}$-plet bulk fermions 
   with $N_f=1$ and $N_f = 2$, respectively. 
We have taken $Q = 5/3$ for the color triplet ${\bf 10}$-plets. 
Note that along the dash-dotted line the relation of $m_0 \geq M_{\rm KK}/2$ is satisfied, 
  and we can also consider a half-periodic boundary condition for this case with $N_f^{\rm HP}=1$.

%%%%%%%%%%%%%%%%%%%%%%%%%%%%%%%%%%%%%%%%%%%%%%%%%%%%
% Fig 5
%%%%%%%%%%%%%%%%%%%%%%%%%%%%%%%%%%%%%%%%%%%%%%%%%%%%%
%\begin{figure}[htbp]
\begin{figure}[t]
  \begin{center}
 \includegraphics[width=130mm]{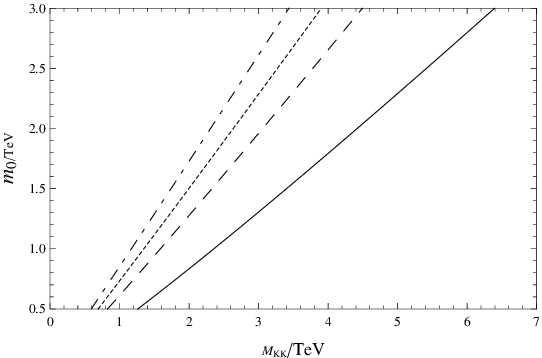}
   \end{center}
\caption{
The relation between $M_{\rm KK}$  and $m_0$ to reproduce the Higgs boson pole mass of $m_H=125.09$ GeV. 
The solid and dashed lines denote the results for the color singlet, ${\bf 10}$-plet bulk fermions 
  with $N_f = 1$ and $N_f = 2$, respectively. 
Here we have taken $Q = 1$.  
The dotted and dash-dotted lines represent the results for the color triplet, ${\bf 10}$-plet bulk fermions 
   with $N_f=1$ and $N_f = 2$  $(N_f^{\rm HP}=1)$, respectively. 
We have taken $Q = 5/3$ for the ${\bf 10}$-plets. 
}
  \label{Fig:MkkVSm0-10}
\end{figure}
%%%%%%%%%%%%%%%%%%%%%%%%%%%%%%%%%%%%%%%%%%%%%%%%%%%%%

%%%%%%%%%%%%%%%%%%%%%%%%%%%%%%%%%%%%%%%%%%%%%%%%%%%%%%
\section{Higgs boson production and decay in GHU model}
%%%%%%%%%%%%%%%%%%%%%%%%%%%%%%%%%%%%%%%%%%%%%%%%%%%%%% 
Through quantum corrections at the one-loop level, the bulk fermions contribute 
   to the Higgs boson production and decay processes and deviate the Higgs boson signal strengths 
   at the LHC experiments from the SM predictions.
In this section, we evaluate the contributions from  the bulk ${\bf 6}$-plet and ${\bf 10}$-plet fermions
   to the Higgs boson production and decay processes at the LHC, 
   and lead to a lower mass bound for the lightest bulk fermion.

%%%%%%%%%%%%%%%%%%%%%%%%%%%%%%%%%%%%%
\subsection{Bulk fermion contributions to the gluon fusion channel} 
%%%%%%%%%%%%%%%%%%%%%%%%%%%%%%%%%%%%%
At the LHC, the Higgs boson is dominantly produced via gluon fusion process 
  with the following dimension five operator between the Higgs boson and di-gluon: 
\bea
{\cal L}_{{\rm eff}} = C_{gg} h G^a_{\mu\nu} G^{a\mu\nu}  ,
\label{dim5g}
\eea
 where $h$ is the SM Higgs boson, and $G^a_{\mu\nu}$ ($a=1-8$) is the gluon field strength.   
The SM contribution to $C_{gg}$ is dominated by top quark 1-loop corrections. 
As a good approximation, we express this contribution by using the Higgs low energy theorem \cite{HLET}, 
\bea
C_{gg}^{{\rm SM top}} \simeq \frac{g_3^2}{32\pi^2 v} 
 b_3^t \frac{\partial}{\partial \log v} \log M_t 
 = \frac{\alpha_s}{12\pi v} ,
\label{SMg}
\eea
  where $\alpha_s=g_3^2/(4 \pi)$, and $b_3^t=2/3$ is a top quark contribution 
  to the beta function coefficient of QCD.

In addition to the SM contribution, we take into account the contributions from 
  the top quark KK modes.  
One might think that the KK mode contributions from the light SM fermions should be taken into account. 
However, they can be neglected compared to those from the top quark KK modes, 
  since the effective coupling of the Higgs boson to di-gluon is generated by the electroweak 
  symmetry breaking and hence proportional to the corresponding SM fermion masses.  
Thus, we only consider the top quark KK mode contribution. 
In GHU scenario, we expect the top quark KK mode spectrum to be 
\bea
  m_{n,t}^{(\pm)} = m_n \pm M_t  ,
 \label{KKtopmass}
\eea
  where $m_n \equiv n M_{\rm KK}$ with an integer $n=1,2,3,\cdots$. 
Using the Higgs low energy theorem, we obtain 
\bea
C_{gg}^{{\rm KKtop}} &\simeq& \frac{\alpha_s}{12\pi v}  
 \sum_{n=1}^\infty \frac{\partial}{\partial \log v} 
 \left[ \log (m_n + M_t) + \log(m_n - M_t) \right] \nonumber \\
&\simeq&  - \frac{\alpha_s}{6\pi v}  \sum_{n=1}^\infty 
 \left( \frac{M_t}{m_n} \right)^2
=
- \frac{\alpha_s}{12\pi v} \times \frac{\pi^2}{3} \left( \frac{M_t}{M_{\rm KK}} \right)^2, 
\label{KKgfusion}
\eea
 where we have used an approximation of $M_t^2 \ll m_n^2$, 
 and $\sum_{n=1}^\infty 1/n^2 = \pi^2/6$. 
As pointed out in Ref.~\cite{MO},  the KK mode contribution to the effective Higgs coupling to di-gluon 
 is destructive to the SM one.

The $N_f$ pairs of color triplet, ${\bf 6}$-plet and ${\bf 10}$-plet bulk fermions 
  have the KK mass spectra as shown in Eqs.~(\ref{6KKspectrum}) and (\ref{10KKspectrum}), respectively. 
Applying the Higgs low energy theorem, 
  their contributions to the Higgs-to-digluon coupling are calculated as 
\bea
C_{gg}^{{\rm KK6}} 
 &\simeq&  F(2m_W) + F(m_W), \\
\label{digluon6}
C_{gg}^{{\rm KK10}} 
 &\simeq& F(3m_W) + F(2m_W) + 2F(m_W), \\
\label{digluon10}
\eea
where the function $F(m_W)$ is given by 
\bea
F(m_W) &=& 2 N_f  \frac{\alpha_{s}}{12\pi v} 
 \sum_{n=-\infty}^\infty \frac{\partial}{\partial \log v} \left[ \log \sqrt{(m_n + m_W)^2+M^2}  \right]\nonumber \\
&\simeq&
2 N_f  \frac{\alpha_{s}}{12\pi v}  \left(\frac{m_W}{M_{\rm KK}} \right)^2
\left[
\frac{1}{c_B^2+(m_W/M_{\rm KK})^2} 
-2  \sum_{n=1}^\infty 
\frac{n^2- c_B^2}{\left(n^2+ c_B^2 \right)^2}  \right]  
\nonumber \\
&=&
2 N_f  \frac{\alpha_{s}}{12\pi v} 
\left(\frac{m_W}{M_{\rm KK}} \right)^2
\left[
\frac{1}{c_B^2+(m_W/M_{\rm KK})^2} 
- \frac{1- (\pi c_B/\sinh[\pi c_B])^2}{c_B^2}
 \right],   
 \label{gfusion-P}
\eea
for the bulk fermion for which a periodic boundary condition is imposed. 
Note that $F(m_W)$ is positive since the zero-mode ($n=0$) contribution dominates 
  over the negative KK-mode contributions. 
When we impose a half-periodic boundary condition, we have \cite{MO2}
\bea
F(m_W) &=&2 
 N_f^{\rm HP}  
\frac{\alpha_{s}}{12\pi v} 
 \sum_{n=-\infty}^\infty \frac{\partial}{\partial \log v} \left[ \log \sqrt{M^2+(m_{n + \frac{1}{2}} + m_W)^2}  \; \right] \nonumber \\
&\simeq&
- 2  N_f^{\rm HP}
 \frac{\alpha_{s}}{6\pi v} \left(\frac{m_W}{M_{\rm KK}} \right)^2
\sum_{n=0}^\infty 
\frac{\left( n+\frac{1}{2} \right)^2- c_B^2}
{\left( \left(n+\frac{1}{2}\right)^2+ c_B^2 \right)^2}  \nonumber\\
&=&
-2 
 N_f^{\rm HP}
\frac{\alpha_{s}}{12 \pi v} \left(\frac{m_W}{M_{\rm KK}} \right)^2
 \frac{\pi^2}{\cosh^2[\pi c_B]}. 
 \label{gfusion-AP}
\eea
Here, $c_B=M/M_{\rm KK}$, and we have used the approximation $m_W^2 \ll M_{\rm KK}^2$. 
 
\begin{figure}[ht]
  \begin{center}
   \includegraphics[width=80mm]{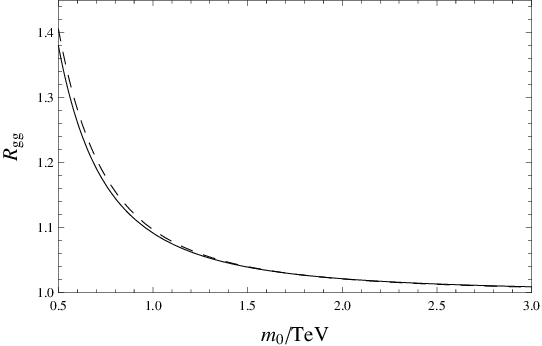}
   \hspace*{5mm}
   \includegraphics[width=80mm]{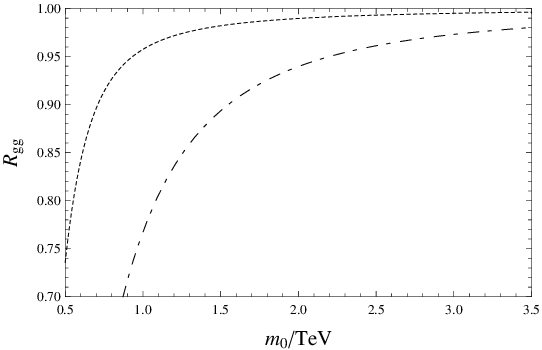}
   \end{center}
  \caption{
The ratio of the Higgs production cross section in our model to the SM one 
  as a function of the lightest bulk fermion mass $m_0$. 
The left panel shows the results for the ${\bf 6}$-plet case 
   corresponding to the right panel of Fig.~\ref{Fig:MkkVSm0}. 
The results for the ${\bf 10}$-plet case, corresponding to the dotted and dash-dotted lines 
   in Fig.~\ref{Fig:MkkVSm0-10},  are depicted in the right panel.  
Here we have considered the periodic boundary condition for the dotted line 
   in Fig.~\ref{Fig:MkkVSm0-10}, while the half-periodic boundary condition 
   for the dash-dotted line in Fig.~\ref{Fig:MkkVSm0-10}.  
The dotted and dash-dotted lines represent the results for the periodic and half-periodic 
   ${\bf 10}$-plet fermions, respectively. 
}
  \label{ggh}
\end{figure}

Now we evaluate the ratio of the Higgs production cross section through the gluon fusion 
   at the LHC to the SM one as 
\bea
R_{gg} &\equiv& 
 \left( 1 + \frac{C_{gg}^{\rm KKtop}+C_{gg}^{\rm KK6/10}}{C_{gg}^{\rm SMtop}} \right)^2 . 
\eea
In the previous section, we have found a relation between $M_{\rm KK}$ and $m_0$ 
  for the color triplet bulk fermions to reproduce the Higgs boson pole mass of $m_H=125.09$ GeV 
  (see Figs.~\ref{Fig:MkkVSm0} and \ref{Fig:MkkVSm0-10}). 
With the relation, we can express the ratio $R_{gg}$ as a function of $m_0$. 
Our results are shown in Fig.~\ref{ggh}. 
The left panel shows the ratio in the presence of the color-triplet, bulk ${\bf 6}$-plet fermions    
  for which the relation between $M_{\rm KK}$ and $m_0$ are shown 
  in the right panel of Fig.~\ref{Fig:MkkVSm0}. 
The solid and dashed lines in the left panel of Fig.~\ref{ggh} correspond to 
  the solid and dashed lines in the right panel of Fig.~\ref{Fig:MkkVSm0}. 
The right panel of Fig.~\ref{ggh} shows the results for the color-triplet, ${\bf 10}$-plet bulk fermions.  
The relations between $M_{\rm KK}$ and $m_0$ for the fermions are shown 
   as the dotted and the dash-dotted lines in Fig.~\ref{Fig:MkkVSm0-10}. 
Note that the relation along the dash-dotted  line satisfies $m_0 \geq M_{\rm KK}/2$, so that 
   we can also apply a half-periodic boundary condition for the ${\bf 10}$-plet fermions with $N_f^{\rm HP}=1$. 
Here we have considered the periodic boundary condition for the dotted line 
   in Fig.~\ref{Fig:MkkVSm0-10}, while the half-periodic boundary condition 
   for the dash-dotted line Fig.~\ref{Fig:MkkVSm0-10}.  
The dotted and dash-dotted lines represent the results for the periodic and half-periodic 
   ${\bf 10}$-plet fermions, respectively.

%%%%%%%%%%%%%%%%%%%%%%%%%%%%
% Fig 6
%%%%%%%%%%%%%%%%%%%%%%%%%%%%
%%%%%%%%%%%%%%%%%%%%%%%%%%%%%%%%%%%%%%%%%%%%%%%%%%%%%

In the presence of the bulk fermions, the Higgs production cross section 
  in the gluon fusion channel is altered from the SM prediction. 
This deviation becomes larger as $m_0$ (or equivalently, $M_{\rm KK}$) is lowered.  
Since the Higgs boson properties measured by the LHC experiments 
  are found to be consistent with the SM predictions \cite{ATL_CMS_Hproperties}, 
  we can find a lower bound for $m_0$ from the LHC results. 
Employing the results from a combined analysis by the ATLAS and the CMS collaborations \cite{ATL_CMS_Hproperties}, 
   $0.89 \leq  R_{gg} \leq 1.19$, 
   we can read off a lower bound of the lightest bulk fermion mass $m_0$ from Fig.~\ref{ggh}.  
Our results are summarized in Table~\ref{table1}. 
The lower bounds are found to be at the TeV scale, so that such exotic colored particles 
  can be tested at the LHC Run-2 with $\sqrt{s}=13-14$ TeV.

%%%%%%%%%%%%%%%%%%%%%%%%%%%%%%%%%%%%%%%%%%
%\begin{table}[htb]
\begin{table}[t]
  \begin{center}
    \begin{tabular}{|c|c|c|c||c|c|} 
\hline
   & BC & $N_f^{(\rm{HP})}$ & $Q$ & $m_0$ (TeV) & $M_{\rm KK}$ (TeV) \\  
\hline \hline
${\bf 6}$-plet & P & $1$ & $4/3$ & $0.703$  & $2.89$  \\
${\bf 6}$-plet & P & $2$ & $4/3$ & $0.728$  & $1.67$  \\
\hline
${\bf 10}$-plet & P   & $1$ & $5/3$ & $ 0.685$  & $0.937$  \\
${\bf 10}$-plet & HP & $1$ & $5/3$ & $1.48$  & $1.71$  \\
\hline 
top quark KK mode &  &  & &  & $1.32 $  \\
\hline
\end{tabular}
  \caption{
The lower bound on the lightest bulk fermion masses and 
  the compactification scales from the ATLAS and CMS combined analysis, 
 $0.89 \leq R_{gg} \leq 1.19$, 
 for the cases with the color-triplet, ${\bf 6}$ or ${\bf 10}$-plet bulk fermion. 
Here, the initials, ``BC'', ``P'' and ``HP'' stand for ``boundary condition", ``periodic" and ``half-periodic", respectively. 
We have also shown in the last row the lower bound on the compactification scale 
   when only the top quark KK modes is taken into account.  
}
  \label{table1}
  \end{center}
\end{table}
%%%%%%%%%%%%%%%%%%%%%%%%%%%%%%%%%%%%%%%%%%

%%%%%%%%%%%%%%%%%%%%%%%%%%%%%%%%%%%%%%%
\subsection{Bulk fermion contributions to $h \to \gamma \gamma$}
%%%%%%%%%%%%%%%%%%%%%%%%%%%%%%%%%%%%%%%
Since the bulk fermions have electric charges, 
  they also contribute to the effective Higgs boson coupling with di-photon
  of the dimension five operator, 
\bea
{\cal L}_{{\rm eff}} = C_{\gamma\gamma} h F_{\mu\nu} F^{\mu\nu},  
\eea
   where $F_{\mu\nu}$ denotes the photon field strength. 
In the SM, this effective coupling is induced by the top quark and $W$-boson loop corrections. 
In addition to the SM contributions, 
  we have contributions from  the KK modes of top quark, $W$-boson, and 
  the ${\bf 6}$-plet or ${\bf 10}$-plet bulk fermion.

We begin with the top quark loop contribution. 
By using the Higgs low energy theorem, we have 
\bea
C_{\gamma\gamma}^{{\rm SMtop}} \simeq 
 \frac{e^2 b_1^t}{32\pi^2 v} \frac{\partial}{\partial \log v} \log m_t 
= \frac{2\alpha_{em}}{9\pi v}, 
\label{SMt2gamma}
\eea
 where $b_1=(4/3) \times (2/3)^2 \times 3 =4/3$ is a top quark contribution 
   to the QED beta function coefficient, and $\alpha_{em}$ is the fine structure constant. 
Corresponding KK top quark contribution is given by 
\bea
C_{\gamma\gamma}^{{\rm KKtop}} &\simeq& \frac{e^2 b_1^t}{32\pi^2 v} 
\sum_{n=1}^\infty \frac{\partial}{\partial \log v} \left[ \log (m_n + M_t)  + \log (m_n - M_t) \right] \nonumber \\
&\simeq& 
-\frac{2\alpha_{em}}{9\pi v} \times \frac{\pi^2}{3} 
 \left(\frac{M_t}{M_{\rm KK}} \right)^2. 
\label{KKt2gamma}
\eea
As the same with the contribution to the effective Higgs coupling with digluon, 
  the KK top quark contribution is destructive to the top quark contribution.

Applying the Higgs low energy theorem, the SM $W$-boson loop contribution is calculated as 
\bea
C_{\gamma\gamma}^W &\simeq& \frac{e^2}{32\pi^2 v} b_1^W 
 \frac{\partial}{\partial \log v} \log m_W = -\frac{7\alpha_{em}}{8\pi v}
\label{SMW2gamma}
\eea
 where $m_W=g_2 v/2$, and $b_1^W=-7$ is a $W$-boson contribution to the QED beta function coefficient. 
Since  $4 m_W^2/m_h^2 \gg 1$ is not well satisfied, this estimate is rough. 
In the following numerical analysis, we use the known loop-function for the $W$-boson loop correction \cite{HLET}.

In our model, the KK mode mass spectrum of the $W$-boson 
 is given by 
\bea
 m_{n,W}^{(\pm)} = m_n \pm m_W,  
\eea  
 so that the contribution from KK $W$-boson loop diagrams
 is found to be 
\bea
C_{\gamma\gamma}^{{\rm KKW}} &=& 
 \frac{e^2}{32\pi^2 v} b_1^W \sum_{n=1}^\infty 
 \frac{\partial}{\partial \log v} 
\left[ \log (m_n + m_W) + \log (m_n - m_W) \right]  \nonumber \\
&\simeq& 
\frac{7\alpha_{em}}{8 \pi v} \frac{\pi^2}{3} 
 \left( \frac{m_W}{M_{\rm KK}} \right)^2. 
\label{KKW2gamma}
\eea
Note again that the KK $W$-boson contribution 
 is destructive to the SM $W$-boson contribution.

Finally, the ${\bf 6}$-plet or ${\bf 10}$-plet loop contributions 
  can be read from the KK-mode mass spectrum 
  in Eq.~(\ref{6KKspectrum}) or (\ref{10KKspectrum})  and the electric charges of each modes: 
\bea
C_{\gamma\gamma}^{{\rm KK6}} 
 &\simeq& \left(Q-\frac{2}{3} \right)^2 \hat{F}(2m_W) + \left(Q + \frac{1}{3} \right)^2 \hat{F}(m_W) \\
\label{diphoton6}
C_{\gamma\gamma}^{{\rm KK10}} 
 &\simeq& (Q-1)^2 \hat{F}(3m_W) + (Q-1)^2 \hat{F}(m_W) + Q^2 \hat{F}(2m_W) + (Q+1)^2 \hat{F}(m_W), 
\label{diphoton10}
\eea
where $Q$ is a $U(1)^\prime$ charge for the ${\bf 6}$ and ${\bf 10}$-plets, and 
\bea
{\hat F}(m_W) \simeq
2 N_f N_c  \frac{\alpha_{em}}{6\pi v} 
\left(\frac{m_W}{M_{\rm KK}} \right)^2
\left[
\frac{1}{c_B^2+(m_W/M_{\rm KK})^2} 
- \frac{1- (\pi c_B/\sinh[\pi c_B])^2}{c_B^2}
 \right],   
 \label{diphoton-P}
\eea
for the periodic bulk fermions, 
while for the half-periodic bulk fermions, we have 
\bea
{\hat F}(m_W) \simeq  -2 
N_f^{\rm HP}
N_c \frac{\alpha_{em}}{6 \pi v} \left(\frac{m_W}{M_{\rm KK}} \right)^2
 \frac{\pi^2}{\cosh^2[\pi c_B]}. 
 \label{diphoton-AP}
\eea
Here, we have used the approximation $m_W^2 \ll M_{\rm KK}^2$. 
Similarly to Eq.~(\ref{gfusion-P}), ${\hat F}(m_W)$ in Eq.~(\ref{diphoton-P}) is positive 
  since the zero-mode contribution dominates.

\begin{figure}[ht]
  \begin{center}
   \includegraphics[width=80mm]{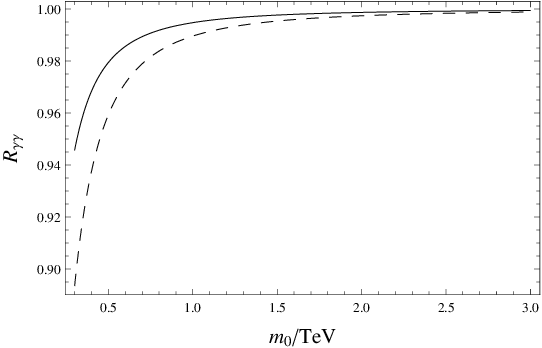}
   \hspace*{5mm}
   \includegraphics[width=80mm]{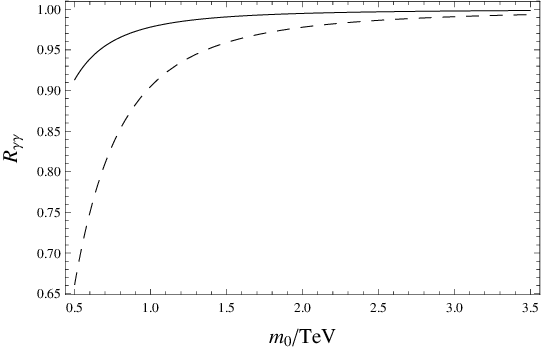}
   \end{center}
  \caption{
The signal strength for the Higgs diphoton decay mode in our model 
   as a function of the lightest bulk fermion mass $m_0$. 
The left panel shows the results for the ${\bf 6}$-plet case 
   corresponding to the left panel of Fig.~\ref{Fig:MkkVSm0}. 
The results for the ${\bf 10}$-plet case, which correspond to 
  the solid and dashed lines in Fig.~\ref{Fig:MkkVSm0-10},  
  are depicted in the right panel.  
}
  \label{Fig:Rgam}
\end{figure}

The ratio of the partial decay width of $h \to \gamma \gamma$ 
  in our model to the SM one is given by
\bea
R_{\gamma\gamma} = \left( 1+ \frac{C_{\gamma\gamma}^{{\rm KKtop}} 
+ C_{\gamma\gamma}^{{\rm KKW}} + C_{\gamma\gamma}^{{\rm KK6/10}}}
{C_{\gamma\gamma}^{{\rm SMtop}} + C_{\gamma\gamma}^W} \right)^2 .
\eea
If the bulk fermions are color-singlet,  they have no effect on the Higgs boson production cross section. 
In this case, the effect of the bulk fermions can be seen in a deviation of 
  the signal strength of the Higgs diphoton decay mode.  
Since the branching fraction of $h \to \gamma \gamma $ is of order 0.1\%,  
  the deviation of the signal strength from the SM prediction 
  is evaluated by $R_{\gamma \gamma}$.   
The relation between $m_0$  and $M_{\rm KK}$ is determined so as to reproduce 
  $m_H=125.09$ GeV, we evaluate $R_{\gamma \gamma}$ as a function of $m_0$ 
  for the color-singlet, ${\bf 6}$-plet and ${\bf 10}$-plet bulk fermions.  
Our results are shown in Fig. ~\ref{Fig:Rgam}. 
The left panel shows the results for the ${\bf 6}$-plet case presented in the left panel of Fig.~\ref{Fig:MkkVSm0}.  
The solid and dashed lines are corresponding to the same types of lines 
   in the left panel of Fig.~\ref{Fig:MkkVSm0}.  
The results for the ${\bf 10}$-plet case, which correspond to 
  the solid and dashed lines in Fig.~\ref{Fig:MkkVSm0-10},  
  are depicted in the right panel.

%%%%%%%%%%%%%%%%%%%%%%%%%%%%
% Fig 7
%%%%%%%%%%%%%%%%%%%%%%%%%%%%
%%%%%%%%%%%%%%%%%%%%%%%%%%%%%%%%%%%%%%%%%%%%%%%%%%%%%

To derive a lower bound on $m_0$, we employ the results of the signal strength  
  from a combined analysis by the ATLAS and the CMS collaborations \cite{ATL_CMS_Hproperties} 
  such as $0.96 \leq  \mu^{\gamma \gamma} \leq 1.33$, 
  which is identified as $R_{\gamma \gamma}$ in our case. 
We then read off a lower bound  on the lightest bulk fermion mass $m_0$ from Fig.~\ref{Fig:Rgam}.  
Our results are summarized in Table~\ref{table2}.

%%%%%%%%%%%%%%%%%%%%%%%%%%%%%%%%%%%%%%%%%%
\begin{table}[htb]
  \begin{center}
    \begin{tabular}{|c|c|c|c||c|c|} 
\hline
   & BC & $N_f$ & $Q$ & $m_0$ (TeV) & $M_{\rm KK}$ (TeV) \\  
\hline \hline
${\bf 6}$-plet & P & $1$ & $2/3$ & $0.353$  & $17.2$  \\
${\bf 6}$-plet & P & $2$ & $2/3$ & $0.504$  & $5.48$  \\
\hline
${\bf 10}$-plet & P & $1$ & $1$ & $0.744$  & $1.81$  \\
${\bf 10}$-plet & P & $2$ & $1$ & $1.52$  & $2.35$  \\
\hline 
\end{tabular}
  \caption{
The lower bound on the lightest bulk fermion masses and 
  the compactification scales from the ATLAS and CMS combined analysis, 
 $0.96 \leq \mu^{\gamma \gamma} \simeq R_{\gamma \gamma} \leq 1.33$, 
 for the cases with the color-singlet, ${\bf 6}$ and ${\bf 10}$-plet bulk fermions. 
}
  \label{table2}
  \end{center}
\end{table}
%%%%%%%%%%%%%%%%%%%%%%%%%%%%%%%%%%%%%%%%%%

%%%%%%%%%%%%%%%%%%%%%%%%%%%%
% Fig 8
%%%%%%%%%%%%%%%%%%%%%%%%%%%%
\begin{figure}[ht]
  \begin{center}
   \includegraphics[width=80mm]{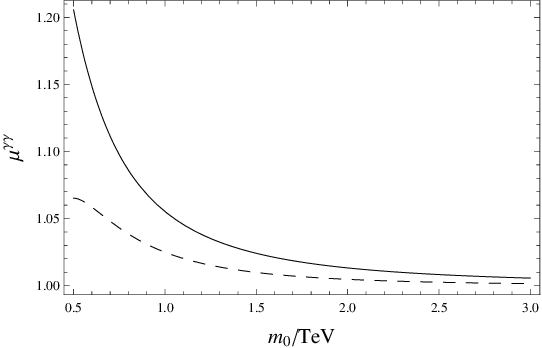}
   \hspace*{5mm}
   \includegraphics[width=80mm]{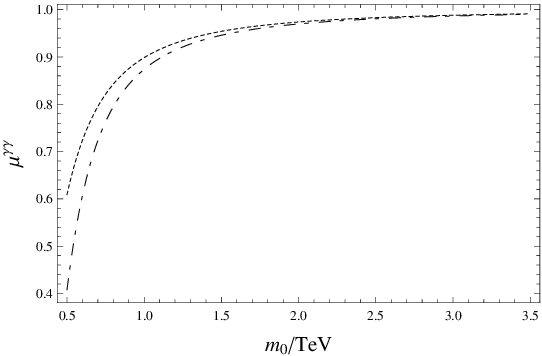}
   \end{center}
  \caption{
The signal strength as a function of the lightest bulk fermion mass $m_0$. 
The left panel shows the results for the ${\bf 6}$-plet case 
   corresponding to the right panel of Fig.~\ref{Fig:MkkVSm0}. 
The results for the ${\bf 10}$-plet case, corresponding to the dotted and dash-dotted lines 
   in Fig.~\ref{Fig:MkkVSm0-10},  are depicted in the right panel.     
Here we have considered the periodic boundary condition for the dotted line 
   in Fig.~\ref{Fig:MkkVSm0-10}, while the half-periodic boundary condition 
   for the dash-dotted line Fig.~\ref{Fig:MkkVSm0-10}.  
The dotted and dash-dotted lines represent the results for the periodic and half-periodic 
   ${\bf 10}$-plet fermions, respectively. 
}
  \label{Fig:Mugam}
\end{figure}
%%%%%%%%%%%%%%%%%%%%%%%%%%%%%%%%%%%%%%%%%%%%%%%%%%%%%

%%%%%%%%%%%%%%%%%%%%%%%%%%%%%%%%%%%%%%%%%%
\begin{table}[htb]
  \begin{center}
    \begin{tabular}{|c|c|c|c||c|c|} 
\hline
   & BC & $   
    N_f^{(\rm{HP})}
   $ & $Q$ & $m_0$ (TeV) & $M_{\rm KK}$ (TeV) \\  
\hline \hline
${\bf 10}$-plet & P    & $1$ &  $5/3$ & $1.61$  & $2.13$  \\
${\bf 10}$-plet & HP & $1$ &  $5/3$ & $1.74$  & $2.01$  \\
\hline 
\end{tabular}
  \caption{
The lower bound on the lightest bulk colored fermion masses and 
  the compactification scales from the ATLAS and CMS combined analysis, 
 $0.96 \leq \mu^{\gamma \gamma} \leq 1.33$.  
We have obtained the lower bound only for the ${\bf 10}$-plet bulk fermions. 
}
  \label{table3}
  \end{center}
\end{table}
%%%%%%%%%%%%%%%%%%%%%%%%%%%%%%%%%%%%%%%%%%

%%%%%%%%%%%%%%%%%%%%%%%%%%%%%%%%%%%%%%%%%%%%%%%
\subsection{Bulk colored fermion contributions to $gg \to h \to \gamma\gamma$}
%%%%%%%%%%%%%%%%%%%%%%%%%%%%%%%%%%%%%%%%%%%%%%%
Finally, we calculate the signal strength of the process 
  $gg \to h \to \gamma \gamma$ in our model. 
The bulk colored fermions contribute to both the effective Higgs coupling to di-gulon and di-photon, 
  and hence alter this signal strength from the SM prediction. 
The signal strength is calculated by     
\bea
\mu^{\gamma \gamma} \simeq  \frac{\sigma(gg \to h \to \gamma \gamma)
}{\sigma(gg \to h \to \gamma \gamma)_{\rm SM}}
=R_{gg} \times R_{\gamma\gamma} . 
\eea
We show our results in Fig.~\ref{Fig:Mugam} for the color-triplet, ${\bf 6}$-plet (left panel) 
   and ${\bf 10}$-plet fermions.  
The left panel shows the results for the ${\bf 6}$-plet case presented in the right panel of Fig.~\ref{Fig:MkkVSm0}.  
The solid and dashed lines are corresponding to the same types of lines 
   in the left panel of Fig.~\ref{Fig:MkkVSm0}.  
The results for the ${\bf 10}$-plet case, which correspond to 
  the dotted and the dash-dotted lines in Fig.~\ref{Fig:MkkVSm0-10},  
  are depicted in the right panel.  
As the same in Fig.~\ref{ggh} the dotted line represents the case with the periodic boundary condition, 
   while the dash-dotted line corresponds to the case with the half-periodic boundary condition.

Employing the constraint, $0.96 \leq  \mu^{\gamma \gamma} \leq 1.33$, 
   from the ATLAS and CMS combined analysis \cite{ATL_CMS_Hproperties},   
   we can find a lower bound on $m_0$ from Fig.~\ref{Fig:Mugam}.  
No lower bound can be obtained from the results on the left panel. 
For the ${\bf 6}$-plet fermions, we may apply the lower bound presented in Table~\ref{table1}.  
When we assign $Q=4/3$ to the ${\bf 6}$-plet fermions, the lightest mode has the color-triplet 
   with an electric charge $2/3$, so that it can generally mix with the SM top quark.  
Through this mixing, once produced dominantly through the gluon fusion process at the LHC, 
   it can decay into the $W$-boson/$Z$-boson/Higgs boson and top/bottom quark 
   through the charged/neutral current.   
The current search for such a vector-like color triplet particle at the LHC has set 
  lower mass limits between $720$ and $920$ GeV at 95\% confidence level \cite{T-search},  
  which are more severe than those listed in Table~\ref{table1}.  
The lower bounds on $m_0$ can be read off from the right panel in Fig.~\ref{Fig:Mugam} 
   and are summarized in Table~\ref{table3}. 
Comparing Tables~\ref{table1} and \ref{table3},  we see that the lower mass bounds 
   from $\mu^{\gamma \gamma}$ for the ${\bf 10}$-plet are more severe than those 
   from $R_{gg}$ and the direct search at the LHC.

%%%%%%%%%%%%%%%%%%%%%%%%%%%%%%%%%
\section{Conclusions}
%%%%%%%%%%%%%%%%%%%%%%%%%%%%%%%%%
Since the discovery of the SM Higgs boson at the LHC,  
   the properties of the Higgs boson have been investigated 
   towards the experimental confirmation of the origin of mass 
   and electroweak symmetry breaking in the SM.     
The LHC Run-2 with the upgrade of the collider energy to 13 TeV 
   is in operation and more data is being accumulated. 
In the near future, the Higgs boson properties such as 
   its mass and decay rates to a variety of modes will be more accurately measured, 
   by which the Higgs sector in the SM may be precisely confirmed 
   or some deviation from the SM framework may be revealed.

The gauge hierarchy problem is one of the most serious problems of the SM, 
   and new physics models have been proposed towards 
   a solution to the problem. 
Such new physics models include new particles whose couplings to the SM Higgs doublet 
   influence the Higgs boson properties. 
For example, in the minimal supersymmetric standard model,  
  there is a correlation between the Higgs boson mass and the mass of 
  sparticles, in particular, scalar top quarks.

In this paper, we have considered the gauge-Higgs unification scenario 
  in 5-dimensional flat space-time,  
  where the SM Higgs doublet is embedded in the 5th spacial component of 
  the gauge field in 5-dimensions.  
Thanks to the gauge symmetry, the quadratic divergence of the Higgs self-energy 
  is forbidden and as a result, the gauge hierarchy problem can be solved.  
The gauge symmetry also forbids the Higgs potential at the tree-level,  
  which is generated through quantum corrections with the breaking 
  of the original bulk gauge symmetry down to the SM one. 
Thus, once the model is defined, the Higgs potential is in principle calculable 
  and the Higgs boson mass can be predicted. 
However, it is highly non-trivial to propose a concrete and complete GHU model 
  which can provide not only a realistic (effective) Higgs potential but also all realistic SM fermion mass matrices.

In analyzing the Higgs boson properties in the context of the GHU scenario,  
   there is a powerful approach in the low energy effective theoretical point of view.  
Independently of details of the GHU models, the Higgs potential must disappear 
   once the bulk gauge symmetry is restored at some high energy, 
   which is identified as the compactification scale 
   through analysis of the effective Higgs potential in a simple GHU model.   
In low energy effective theory where all the KK modes are decoupled, 
   this general feature of the GHU scenario yields the so-called gauge-Higgs condition, 
   namely,  the Higgs quartic coupling is set to be zero at the compactification scale. 
Therefore, under the assumption that the electroweak symmetry breaking is correctly achieved, 
   the Higgs boson mass can be calculated by extrapolating the vanishing Higgs quartic coupling 
   at the compactification scale towards low energies.

For a simple GHU model based on the bulk gauge group SU(3)$\times$U(1)$^\prime$, 
   we have analyzed the RG equations with the gauge-Higgs boundary condition 
   to reproduce the observed Higgs boson pole mass of $m_H=125.09$ GeV.  
If we assume only the SM particle contents below the compactification scale, 
   the Higgs boson mass is realized by $M_{\rm KK} \simeq 10^{10}$ GeV.   
We have introduced bulk fermions with a bulk mass 
   and imposed a periodic or half-periodic boundary condition.  
For concreteness, we have considered color-singlet/triplet, ${\bf 6}$ and ${\bf 10}$-plets 
   of SU(3) with a U(1)$^\prime$ charge $Q$. 
Once the fermion representation is fixed, we have found a unique relation 
   between the compactification scale and the bulk mass so as to reproduce $m_H=125.09$ GeV. 
In the presence of the bulk fermions with the TeV scale mass, 
  we have found $M_{\rm KK} \ll 10^{10}$ GeV, which is desired in the naturalness point of view.

We have also investigated the effect of the bulk fermions on the Higgs boson phenomenology. 
The bulk fermions contribute to the effective Higgs couplings to di-gluon and/or di-photon 
    through quantum correction at the one-loop level and as a result,  
    the Higgs boson production and decay rates at the LHC can be altered from the SM predictions. 
We have employed the current LHC data, which are consistent with the SM predictions, 
    and derived the lower mass bound on the lightest KK mode fermion. 
More precise measurements of the Higgs production and decay rates in the future 
    can indirectly test the existence of the bulk fermions.

As pointed out in the second paper on Ref.~\cite{MO2}, the bulk fermions, 
   if their masses lie in the TeV scale, can also be tested directly at the LHC. 
A bulk fermion multiplet includes many fermions with a variety of electric charges, 
  and  the mass splittings among the fermions are predicted after the electroweak symmetry breaking. 
Hence, a heavy fermion, once produced at the LHC, causes cascade decays to 
   lighter mass eigenstates and the weak gauge bosons or the Higgs boson, 
   which end up with the lightest KK mode fermion.  
The lightest KK mode fermion can be a dark matter candidate if it is electrically neutral and stable,  
  or decays to the SM fermions through a mixing with them. 
The existence of the variety of fermion mass eigenstates and their cascade decay 
    at the LHC are a characteristic feature of our GHU model.  
Search for the KK mode fermions at the LHC Run-2 is an interesting topic and we leave it for future work.

%%%%%%%%%%%%%%%
\section*{Acknowledgments}
%%%%%%%%%%%%%%% 
We would like to thank Satomi Okada for her useful comments on the manuscript. 
N.O. would like to thank KEK Theory Center for hospitality during his visit. 
This work is supported in part by the United States Department of Energy Grant, No.~DE-SC0013680.

%%%%%%%%%%%%%%%

\end{document}